\documentclass[
]{ceurart}

\usepackage{listings}
\begin{document}

\copyrightyear{2026}
\copyrightclause{Copyright for this paper by its authors.
  Use permitted under Creative Commons License Attribution 4.0
  International (CC BY 4.0).}

\conference{ISWC'26: Posters and Demos,
  October 25--29, 2026, Bari, IT}

\title{COCI: Conference Organisers and Content Identifier}


\author[1]{Angelo Salatino}[%
orcid=0000-0002-4763-3943,
email=angelo.salatino@open.ac.uk,
]
\cormark[1]
\address[1]{Knowledge Media Institute, The Open University, UK}

\author[1]{Francesco Osborne}[%
orcid=0000-0001-6557-3131,
email=francesco.osborne@open.ac.uk,
]

\author[2]{Alexis Vizcaino}[%
orcid=0000-0003-4364-284X,
email=alexis.vizcaino@springernature.com
]
\address[2]{Springer Nature, Germany}

\author[2]{Aliaksandr Birukou}[%
orcid=0000-0002-4925-9131,
email=aliaksandr.birukou@springer.com
]

\author[1]{Enrico Motta}[%
orcid=0000-0003-0015-1952,
email=enrico.motta@open.ac.uk
]

\cortext[1]{Corresponding author.}

\begin{abstract}
Despite the critical role of grey literature in scholarly communication, artefacts such as Calls for Papers (CfPs) remain largely isolated from modern Scholarly Knowledge Graphs. The unstructured and highly heterogeneous nature of these documents has traditionally hindered their large-scale processing. In this demo paper, we present the Conference Organisers and Content Identifier (COCI), an AI-based framework designed to extract fine-grained, structured metadata from raw CfP texts. COCI employs a multi-stage pipeline that combines Large Language Models (LLMs) with semantic mapping techniques to integrate extracted entities with established knowledge bases, including OpenAlex, DBLP, TIB ConfIDent, and the AIDA Dashboard. By disambiguating authors and semantically aligning topics and conference series, COCI bridges the gap between informal scholarly dissemination and structured Semantic Web resources, laying the foundation for systematic analysis of non-publisher-based academic events.
\end{abstract}


\maketitle

\section{Introduction}

The majority of Metascience and Scientometric analyses traditionally focus on extracting insights from formal published literature, such as journal issues or conference proceedings. While these studies have yielded a significant understanding of how science evolves~\cite{klavans2006quantitative,angioni2022leveraging}, how research teams are composed~\cite{salatino2026}, and how to detect emerging trends~\cite{salatino2018augur}, a vital dimension of research remains largely overlooked: grey literature~\cite{kousha2022high}. Scholarly communication has increasingly expanded beyond formal publishing channels to include various forms of grey literature, such as preprints, technical reports, policy documents, and Calls for Papers (CfPs)~\cite{adams2017shades, Osayande2012}. These artefacts are vital to the community, as they capture emerging research directions well before they reach formal publication channels. 

However, the informal nature of such material introduces varying degrees of challenges for systematic tracking. While some formats like preprints and theses have increasingly adopted standard repositories and DOI indexing, other crucial artefacts remain entirely elusive~\cite{Paez2017,schopfel2010grey}. Calls for Papers perfectly embody the extreme end of these challenges. They completely lack traditional bibliographic control, bypass standard publishing pipelines, and are frequently distributed through ephemeral, informal channels such as mailing lists or static web pages~\cite{bonato2018searching}. Consequently, their highly heterogeneous formats and lack of structure make them exceedingly difficult to analyse at scale.

The emergence of Large Language Models (LLMs) offers a pivotal opportunity to address the technological limitations that previously hindered the harvesting and analysis of CfPs. In this demo paper, we introduce the Conference Organisers and Content Identifier (COCI), an AI-based framework designed to automate the extraction of granular metadata from unstructured CfP texts. COCI processes these raw documents to identify and structure key information, including conference series, geographic locations, comprehensive lists of organisers with their specific roles and affiliations, and topics of interest.

In practice, COCI employs robust semantic alignment strategies to map the resulting entities to external knowledge bases like OpenAlex~\cite{priem2022openalex}, DBLP, TIB ConfIDent~\cite{hagemann2020confident}, and the AIDA Dashboard~\cite{angioni2022aida}. Crucially for the Semantic Web community, this process serves as a vital stepping stone for expanding the depth and richness of Scientific Knowledge Graphs (SKGs)~\cite{salatino2021detection}. Consequently, this unified ecosystem offers metascientists a foundational tool to track science at its earliest embryonic stage, opening new avenues for deeper analyses on the health, scope, and quality of academic conferences and the researchers who organise them.

The code for COCI is available on \url{https://github.com/angelosalatino/oc-conf-detection/releases/tag/v1.2.0}, whereas a video that demonstrates the tool is available on \url{https://youtu.be/hudkTlJ6S_E}.

\begin{figure}[!h]
    \centering
    \includegraphics[width=1\linewidth]{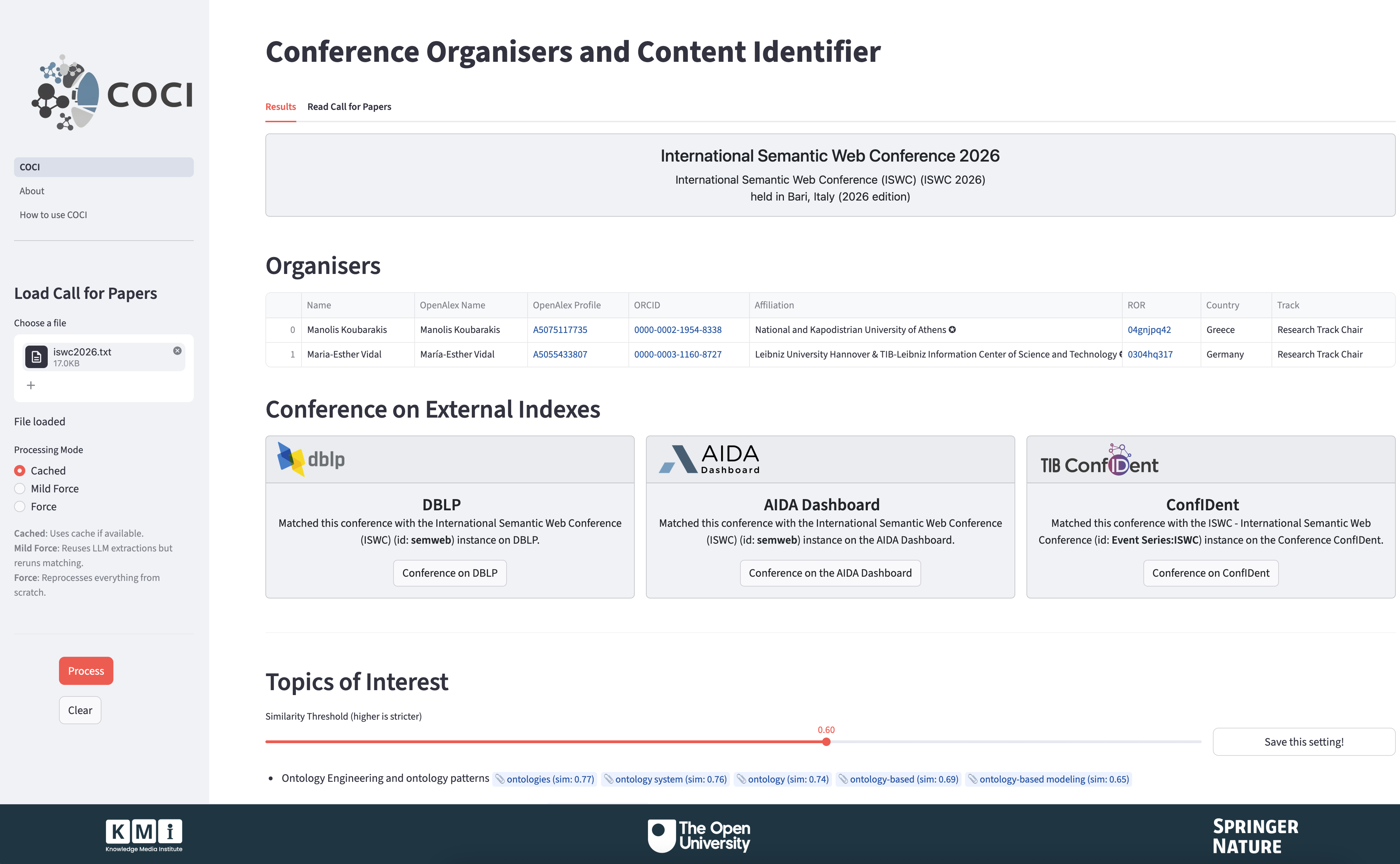}
    \caption{Graphical User Interface of COCI.}
    \label{fig:gui}
\end{figure}

\section{The COCI Application}

COCI is a framework designed to process raw, unstructured text from Call for Papers for scientific events, such as ISWC 2026, and transform it into a rich, structured representation. This output includes essential metadata about the conference (such as the conference series, edition, year, and geographical location), alongside a comprehensive list of the organisers. Crucially, for each organiser, the system extracts their specific track or role and their affiliations, subsequently enriching these profiles with persistent identifiers including OpenAlex IDs, ORCIDs, and Research Organization Registry (ROR) IDs. The application also extracts and structures the event's topics of interest.

To illustrate the system's capabilities, Figure~\ref{fig:gui} displays the application interface showcasing the analysis of the ISWC 2026 Research Track Call for Papers. The top section of the dashboard provides the core conference metadata. Below this, the interface presents the extracted organisers (for example, highlighting \href{https://openalex.org/authors/A5075117735}{Manolis Koubarakis} and \href{https://openalex.org/authors/A5055433807}{María-Esther Vidal}) alongside their affiliations and matched scholarly identifiers. Finally, the bottom section of the display lists the extracted topics of interest, demonstrating how COCI semantically maps these ad-hoc keywords directly to the established OpenAlex topic taxonomy.

The potential of this application extends far beyond automated data entry. By structuring grey literature and mapping it to external knowledge graphs, COCI provides a scalable platform to assess the health and scientific quality of academic events. Structuring the committee lists allows the community to evaluate whether a conference is organised by a diverse and appropriately senior team, and whether it maintains a well-defined scope. Furthermore, this structured representation opens up pathways to connect with external databases, to investigate whether organisers have previously engaged in academic misconduct. These include Retraction Watch Database\footnote{Retraction Watch Database --- \url{https://retractiondatabase.org/RetractionSearch.aspx?}} to check for retractions, and PubPeer\footnote{PubPeer --- \url{https://pubpeer.com/}} to identify if the organisers' publications have been flagged for questionable practices. Ultimately, COCI provides the vital infrastructure needed for a formal recognition platform, enabling a system that grants researchers formal, quantifiable credit for their community contributions, analogous to the model established by Publons for peer review.

\section{Architecture}

\begin{figure}
    \centering
    \includegraphics[width=1\linewidth]{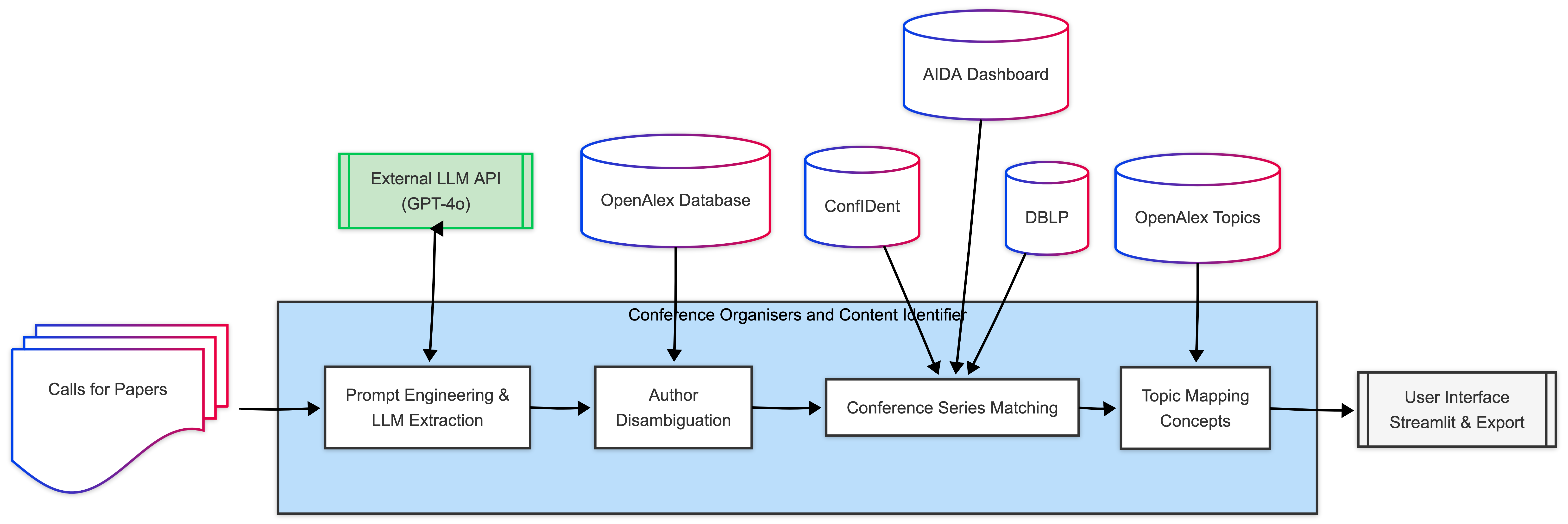}
    \caption{Architecture of COCI}
    \label{fig:architecture}
\end{figure}

The workflow starts with the raw call for papers, where prompt engineering and LLM extraction are first used to structure the unstructured text, as shown in Figure~\ref{fig:architecture}. Then, the pipeline routes the data through a series of dedicated entity matching and semantic validation modules. Specifically, the system performs author disambiguation to link individuals to OpenAlex, conference series matching to align the event with resources like DBLP and the AIDA Dashboard, and topic mapping to ground ad-hoc keywords into controlled conceptual taxonomies. 

\subsection{Prompt Engineering and Extraction}
The extraction process begins by querying the GPT-4o model using a refined prompt template designed to handle the various nuances of the CfPs. This template was fine-tuned through an iterative process involving more than 40 CfP files, ensuring it can accommodate diverse formats across different academic fields. The model parses the raw text to identify key elements, including the event name, acronym, series, year, location, topics of interest, and the organising committee. To guarantee machine-readability, a strict JSON schema is enforced during the API request. The prompt is publicly available at \url{https://github.com/angelosalatino/oc-conf-detection#prompt}.

\subsection{Author and Entity Disambiguation}
To integrate individuals into global scholarly networks, extracted organisers are matched against the OpenAlex database\footnote{OpenAlex --- \url{https://openalex.org/}}. Because author disambiguation is an ongoing challenge in bibliographic databases, we developed an algorithm that combines multiple strategies, including matching author names and institutions while accounting for minor syntactic variations. Specifically, the system prioritises precision by locating the organiser's institution first, mitigating potential LLM hallucinations where identical affiliations might be erroneously repeated. When necessary, COCI falls back on name-only searches, employing heuristics based on publication counts and Levenshtein string similarity to resolve ambiguities. This ultimately enriches the data with persistent identifiers such as ORCIDs\footnote{ORCID --- \url{https://orcid.org/}} and Research Organization Registry\footnote{Research Organization Registry --- \url{https://ror.org/}} (ROR) IDs.

\subsection{Conference Series Alignment}
To accurately situate events within the academic landscape and enhance semantic interoperability, extracted conference series names are matched against three primary scholarly databases: DBLP\footnote{DBLP --- \url{https://dblp.org/}}, the AIDA Dashboard\footnote{AIDA Dashboard --- \url{https://w3id.org/aida/dashboard}}~\cite{angioni2022leveraging}, and TIB ConfIDent\footnote{TIB ConfIDent -- \url{https://www.confident-conference.org/index.php/Main_Page}}~\cite{hagemann2020confident}. This alignment employs a dual-layer approach: initial semantic search is conducted by converting the extracted text into 384-dimensional dense vector embeddings using the \textit{all-MiniLM-L6-v2} SentenceTransformers model\footnote{Sentence Transformers --- \url{https://huggingface.co/sentence-transformers/all-MiniLM-L6-v2}}~\cite{reimers-2019-sentence-bert}. These embeddings are compared against a pre-computed index of all events within the external knowledge bases, and the resulting matches are then rigorously validated using Levenshtein string similarity. This hybrid strategy effectively avoids the false positives common to purely vector-based retrieval. Furthermore, the system leverages cross-referencing capabilities, such that if a match is found in DBLP, internal mappings automatically retrieve the corresponding identifiers for AIDA and ConfIDent, producing a cohesive linked dataset suitable for scientific KGs.

\subsection{Semantic Topic Mapping}
As a final step, COCI maps unstructured, custom topics of interest to controlled vocabularies, specifically the OpenAlex taxonomy\footnote{OpenAlex Topics --- \url{https://help.openalex.org/hc/en-us/articles/24736129405719-Topics}}. Raw extracted research topics are converted into dense vector embeddings using the same sentence transformer model detailed in the previous section. These vectors are then compared against an index of pre-computed OpenAlex concept embeddings using cosine similarity. By default, the system applies a similarity threshold of 0.6 to identify semantically equivalent terms, though the user interface allows this value to be adjusted. This approach bypasses the fragility of literal string matching and rigid syntactic comparison, ensuring that minor variations (such as paraphrasing or hyphenation) are seamlessly mapped to formal research concepts.

\subsection{Visualisation}
COCI presents the processed data through a web interface built with Streamlit (Figure~\ref{fig:gui}), featuring a sidebar for loading the CfP and a main panel that outlines the conference's metadata, organisers, and external records.

The main panel first details core metadata: the conference series, edition, location, year, and acronym. It then presents an interactive table of the organizing committee that maps extracted names to OpenAlex profiles, detailing specific roles, ORCIDs, and ROR-linked affiliations. For deeper context, the system displays links to external records (DBLP, AIDA, and ConfIDent) alongside thematic topics mapped to the OpenAlex taxonomy (highlighted in blue). Lastly, the entire enriched dataset can be exported as a Microsoft Excel file for offline analysis.

\section{Evaluation}
To assess the technical performance and generalisability of COCI across disciplines, we processed and manually evaluated 40 CfPs spanning fields such as Computer Science, Engineering, Scientometrics, and Materials Science. The evaluation demonstrated that the framework successfully processed all CfPs and extracted their metadata, though it highlighted inherent challenges in processing unstructured grey literature.

Some anomalies relevant to knowledge integration and enrichment occurred when organisers were incorrectly mapped to an OpenAlex author profile under a different name. Investigation revealed this was not a failure of COCI's alignment algorithm, but rather an existing error within the OpenAlex database itself, which listed the organiser's name as an alternative alias for another researcher. This instance underscores the fact that the reliability of semantic extraction is sometimes contingent upon the quality of the external Linked Data. Furthermore, the system includes heuristics to tackle LLM hallucinations, such as rejecting affiliation data if the number of unique organisers vastly exceeds the number of unique organisations.

\section{Conclusions}

In this demo paper, we introduced COCI, an AI-driven framework that addresses the longstanding challenge of processing unstructured grey literature. By automating entity extraction and applying rigorous semantic mapping to resources such as DBLP, TIB ConfIDent, the AIDA Dashboard, and OpenAlex, COCI transforms heterogeneous CfP texts into standardised, structured metadata suitable for integration into Scientific Knowledge Graphs. This structured data creates new opportunities to identify researchers who frequently assume community duties, and to assess conference quality based on its scope, the seniority of the organising team, whether any of them engaged in misconduct, and ultimately provide the vital infrastructure for a formal recognition platform. By interconnecting these isolated, non-publisher events, we envision this enriched dataset enabling a system that grants researchers formal, quantifiable credit for their community contributions.

As future work, we intend to develop a continuous crawling engine to harvest new calls for papers from established registries (e.g., WikiCFP) and professional mailing lists. We will also focus on implementing systematic methods for longitudinal analyses of scholarly events, allowing the community to track emerging paradigms and identify disciplinary intersections or collaboration networks at their inception. By enriching existing knowledge graphs with comprehensive event and organiser metadata, COCI paves the way for novel scientometric applications and a deeper integration of informal scholarly communications into the broader research landscape.

\begin{acknowledgments}
We would like to thank Springer Nature for funding this research.
\end{acknowledgments}

\section*{Declaration on Generative AI}
 During the preparation of this work, the author(s) used Gemini in order to: Grammar and spelling check. After using these tool(s)/service(s), the author(s) reviewed and edited the content as needed and take(s) full responsibility for the publication’s content. 

\bibliography{mybib}

\end{document}